\documentstyle[prc,aps,preprint]{revtex}

\begin{document}
\draft
\tighten

\title{Hamiltonian derivation of a detailed fluctuation theorem}
\author{C. Jarzynski}
\address{Theoretical Division, T-13, MS B288 \\
         Los Alamos National Laboratory \\
         Los Alamos, NM 87545 \\
         {\tt chrisj@lanl.gov}}
%\date{\today}

\maketitle

\begin{abstract}
We analyze the microscopic evolution of a system 
undergoing a far-from-equilibrium thermodynamic process.
Explicitly accounting for the degrees of freedom of
participating heat reservoirs, we derive a hybrid result,
similar in form to both the fluctuation theorem, 
and a statement of detailed balance.
We relate this result to the steady-state 
fluctuation theorem, and to a free energy
relation valid far from equilibrium.
\end{abstract}

\pacs{
      LAUR-99-2903}

\section{Introduction}
\label{sec:intro}

The {\it fluctuation theorem} refers collectively 
to a number of theoretical results in the field of 
nonequilibrium statistical mechanics,
which are striking because they are valid {\it far}
from thermal equilibrium.
Following the original numerical discovery by Evans, Cohen, and
Morriss\cite{ecm}, a {\it transient} fluctuation theorem
(applicable to systems driven away from an initial state
of equilibrium), 
and a {\it steady-state} fluctuation theorem (for systems
in a nonequilibrium steady state), were derived by 
Evans and Searles\cite{es}, and by Gallavotti and Cohen\cite{gc},
respectively, for systems evolving under deterministic
but non-Hamiltonian equations of motion.
These results have stimulated considerable
research\cite{g1,c1,bgg,g2,g3,bcl,r,aes,se2,se3,gc2,k,ls,se1,gav2,maes},
in which the fluctuation theorem has been generalized
(in particular to stochastic evolution) and related
to linear response theory, specific examples have
been studied, and the relation between the transient
and steady-state fluctuation theorems has been discussed.

The steady-state version of the fluctuation theorem 
can be written as:
\begin{equation}
\label{eq:ft}
\lim_{\tau\rightarrow\infty}
{1\over\tau}\ln
{p_\tau(+\overline\sigma)\over
p_\tau(-\overline\sigma)} = 
{\overline\sigma\over k_B},
\end{equation}
where $k_B$ is the Boltzmann constant, and
$p_\tau(\overline\sigma)$ is the probability 
distribution of observing an average entropy production 
rate $\overline\sigma$ over
a time interval of length $\tau$.
The distribution is defined with respect to an
ensemble of trajectory segments of duration
$\tau$, sampled while the system in question evolves
in a nonequilibrium steady state.

Physically, maintaining a system in a nonequilibrium 
steady state requires the participation of one or more 
heat reservoirs, for instance to absorb 
the heat generated by shear forces, or to maintain
boundaries of the system at different temperatures.
Derivations of the fluctuation theorem which have
appeared in the literature (whether pertaining to
systems driven away from equilibrium, or to those
in a nonequilbrium steady state) have discussed a variety
of thermostating schemes, both 
deterministic\cite{ecm,es,gc,g1,c1,bgg,g2,g3,bcl,r,aes,se2,se3,gc2}
and stochastic\cite{k,ls,se1,gav2,maes},
to model the presence of reservoirs.
Many of these schemes originated as numerical
strategies for simulating the microscopic evolution of
a system in thermal contact with a heat reservoir,
without simulating the huge number of
degrees of freedom making up the reservoir itself.
The very fact that the fluctuation theorem seems to
be independent of the thermostating scheme lends it
considerable support.
Indeed, Maes\cite{maes} has argued on quite general
grounds that the fluctuation theorem can be
understood as a Gibbs property of space-time histories;
see also Ref.\cite{mrv} for illustrative examples.

Recently, Crooks\cite{gav2} has shown that the
fluctuation theorem is closely related to another set 
of results\cite{cj1,cj2,gavin1,rn,acta,cd,hatano} 
-- also valid far from equilibrium --
which relate the free energy difference between two
equilibrium states of a system, to the external 
work performed on the system during a {\it non}equilibrium 
process from one state to the other.

The purpose of the present paper is to derive a
result similar to the fluctuation theorem,
{\it by explicitly including the 
degrees of freedom of heat reservoirs in the analysis}, and 
assuming Hamiltonian evolution at the microscopic level.
This approach corresponds
closely to the situation present in a laboratory
experiment, where the ``thermostating'' is precisely
the result of interactions between the system and
the innumerable degrees of freedom which constitute 
its environment.
We will argue that, when all microscopic
degrees of freedom are taken into
account, then there emerges a ``detailed fluctuation theorem''
(Eq.\ref{eq:central} below),
valid for finite times $\tau$, 
and expressed without reference to steady states.

A Hamiltonian treatment of nonequilibrium processes,
similar in spirit to that taken here, was carried out by
Bochkov and Kuzovlev\cite{bk1,bk2} (though with less
emphasis on a distinction between the
``system of interest'' and its ``environment'').
The central result derived below, however, is new,
as is the connection to the fluctuation theorem.
More recently, Eckmann, Pillet, and Rey-Bellet\cite{epr}
have introduced and studied an exactly solvable,
Hamiltonian model of a system (a chain of nonlinear
oscillators) coupled to two heat reservoirs at
different temperatures.
It would be very interesting to establish the
precise relation between the
results which they obtain for their model -- especially in
connection with the nonequilibrium steady state --
and the approach taken in Section \ref{sec:ft} of the
present paper.

In the following two sections, the central result
is stated and derived.
While this result does not
explicitly refer to a nonequilibrium steady state,
we argue in Section \ref{sec:ft} that, under
appropriate circumstances, it leads to the steady-state
fluctuation theorem.
In Section \ref{sec:fe} we show that the nonequilibrium
free energy relation
of Refs.\cite{cj1,cj2,gavin1,rn,acta,cd,hatano} also
follows from the central result of the present paper.
We end with a discussion in Section \ref{sec:disc}.

\section{Statement of central result}
\label{sec:central}

Suppose we have the following ingredients at our disposal:
\begin{enumerate}
\item{}
a finite, classical {\it system of interest}, $\psi$,
\item{}
a number of {\it heat reservoirs}, $\theta_1,\theta_2,\cdots,\theta_N$,
\item{}
and, possibly,
a {\it work parameter}, $\lambda$.
\end{enumerate}
The work parameter is some degree of freedom
which we control externally, for instance an external
field, and which interacts 
directly with $\psi$ (but not with the reservoirs).
The reservoirs are also finite, classical systems,
prepared ahead of time at various
temperatures.
We suppose that we can establish or break thermal
contact between our system of interest and any
of the reservoirs, as we choose.
Finally, we assume that, at the most fundamental level of
description, the system and reservoirs are composed of
a large number of microscopic degrees of freedom,
and that the collection of these
obeys Hamiltonian evolution.

Our ability to directly manipulate $\lambda$, and
to make or break contact with
the reservoirs, allows us to subject our
system of interest to a variety of thermodynamic
processes.
We will take the word {\it process}
to be synonymous with an explicit prescription spelling out
``what we do to the system'' at the macroscopic
level, using the tools at our disposal
(the work parameter and heat reservoirs).
More precisely, a process $\Pi$ is defined by a
set of instructions specifying:
\begin{enumerate}
\item{}
the temperatures ($T_1,\cdots,T_N$) at which to
prepare the reservoirs,
\item{}
when to establish and/or break thermal contact between
$\psi$ and any of the $\theta$'s, and
\item{}
the time-dependence of the work parameter, $\lambda(t)$.
\end{enumerate}
We will refer to items 2 and 3 as the {\it protocol}.
We will restrict ourselves to processes occurring over
a finite interval of time, $[0,\tau]$, and
will say that a process is {\it static} if
the work parameter and thermal contacts
are constant over the course of the process.
Note that a process is defined without reference
to the preparation of the system of interest itself.

Having introduced the notion of a process to specify
``what we do to $\psi$'' at the macroscopic level, 
let us now turn our attention to the microscopic response
of the system of interest and reservoirs.
A complete description of this response is provided by
a trajectory $\Gamma(t)$, detailing the (Hamiltonian) 
evolution of all participating degrees of freedom.
We will, however, be interested in a less complete
description, consisting of:
the microscopic history of $\psi$ itself,
and the net entropy generated, $\Delta S$, over the
course of the process.
By the former, we mean a trajectory ${\bf z}(t)$
specifying the evolution of the microstate of
the system of interest (the position and momentum of each 
constituent degree of freedom), from $t=0$ to $t=\tau$.
By {\it entropy generated}, we mean the quantity
\begin{equation}
\label{eq:ent_gen}
\Delta S \equiv -\sum_{n=1}^N
{\Delta Q_n\over T_n},
\end{equation}
where $\Delta Q_n$ denotes the net heat absorbed by
$\psi$ from the $n$'th reservoir, over the course of
the process.
We justify the nomenclature by noting that, at the macroscopic
level of description, $-\Delta Q_n/T_n$ corresponds to
the net change in the entropy of the $n$'th reservoir.
Thus, $\Delta S$ can be viewed as the change in the
entropy of the {\it environment} of $\psi$
(the collection of reservoirs), which we abbreviate to
``entropy generated''.
(See also the definition of the rate of entropy production
introduced in Ref.\cite{epr}.)

Note that both ${\bf z}(t)$ and
$\Delta S$ can be obtained from the complete 
microscopic description, $\Gamma(t)$:
the former, by projecting out the reservoir degrees
of freedom;
the latter, by using the initial and final conditions
$\Gamma(0)$ and $\Gamma(\tau)$ to compute the
net change in the internal energy of each reservoir.
(See Eq.\ref{eq:defdq} below, and commentary in
Section \ref{sec:disc}).

Because the system of interest interacts with the 
reservoirs, the microscopic evolution of $\psi$ 
itself is not deterministic.
Rather, an initial microstate ${\bf z}_A$
determines a {\it statistical ensemble} of possible
realizations, each characterized by a particular 
history ${\bf z}(t)$, and a particular
value of entropy generated $\Delta S$.
This is the ensemble of realizations which we would obtain
by endlessly repeating the same process, always initializing
$\psi$ in the microstate ${\bf z}_A$;
the difference from one realization to the next
arises solely from the different initial conditions
sampled for the reservoirs.
From this ensemble, let us
imagine constructing the statistic
\begin{equation}
P({\bf z}_B,\Delta S|{\bf z}_A),
\end{equation}
which is 
the joint probability distribution of obtaining
a final microstate ${\bf z}(\tau)={\bf z}_B$, 
and an entropy generated $\Delta S$, conditional
on the initial microstate ${\bf z}(0)={\bf z}_A$.
This joint, conditional probability distribution $P$
will be the object of central interest in this paper.
This is admittedly a somewhat peculiar quantity to investigate.
No attempt will be made to motivate our interest
in this statistic
other than that ``it works'', in the sense that
consideration of $P({\bf z}_B,\Delta S\vert{\bf z}_A)$
leads to the neat result expressed by Eq.\ref{eq:central}
below.

Let us now introduce a final piece of notation.
For an arbitrary process $\Pi^+$, let
its {\it time-reversed} counterpart, $\Pi^-$, 
denote the process obtained by using the same set of 
reservoir temperatures, but carrying out the protocol of
$\Pi^+$ in reverse order (reversing the time-dependence 
of both the work parameter and the thermal contacts 
established and broken).
We will, quite arbitrarily, refer to $\Pi^+$ as the 
``forward'' process and $\Pi^-$ as the ``reverse'' 
process.
When discussing the conditional probability
$P({\bf z}_B,\Delta S|{\bf z}_A)$, computed for
two processes $\Pi^+$ and $\Pi^-$ related by
time-reversal, the notation $P_+$ and $P_-$ 
is used to distinguish between the two cases.

The central result of this paper then asserts that
the probability distributions $P_+$ and $P_-$
satisfy the following relation:
\begin{equation}
\label{eq:central}
{P_+({\bf z}_B,+\Delta S\vert {\bf z}_A) \over
 P_-({\bf z}_A^*,-\Delta S\vert {\bf z}_B^*) } =
 \exp (\Delta S/k_B),
\end{equation}
where the asterisk ($^*$) denotes a reversal of momenta:
$({\bf q},{\bf p})^* = ({\bf q},-{\bf p})$.
To obtain some intuition for what this result 
says, imagine filming the evolution of the system,
work parameter, and reservoirs during one realization 
of the process $\Pi^+$, as the microstate
of $\psi$ evolves from ${\bf z}_A$ to ${\bf z}_B$ 
and the entropy generated is $\Delta S$.
Now imagine running the film backward;
you will then see a realization of the process $\Pi^-$,
with $\psi$ evolving from ${\bf z}_B^*$ to
${\bf z}_A^*$, and entropy generation $-\Delta S$.
Eq.\ref{eq:central} thus relates the conditional
probability of observing one set of events
(${\bf z}_A\rightarrow{\bf z}_B$, $+\Delta S$)
during a given process, to that of observing the
time-reversal of those events
(${\bf z}_B^*\rightarrow{\bf z}_A^*$, $-\Delta S$)
during the time-reversed process:
it states that the ratio of these two
probabilities is just the exponent of the entropy
generated, $\Delta S$, in units of $k_B$.
For a static thermodynamic process, we have $\Pi^+=\Pi^-$, 
and therefore we can drop the subscripts $+$ and $-$ 
appearing in Eq.\ref{eq:central}.

Eq.\ref{eq:central} is a hybrid result,
akin both to the fluctuation theorem 
(through dependence on $\Delta S$),
and to a statement of detailed balance
(because of the appearance of the initial and final 
microstates of $\psi$);
for this reason we refer to it as a
{\it detailed fluctuation theorem}.

Note that if $\Delta S>0$, then the conditional 
probability appearing in the numerator is greater than that 
in the denominator; if $\Delta S<0$, the opposite is true.
This makes intuitive sense:
of the two scenarios, the one
which remains obedient to the second law by generating
positive entropy is more likely than
its disobedient twin, by a factor
exponential in the entropy generated.

The proof of Eq.\ref{eq:central} will follow directly 
from the assumption that evolution in the 
full phase space (including the degrees of freedom 
of $\psi,\theta_1,\cdots,\theta_N$)
is deterministic and Hamiltonian.
For a process $\Pi$,
the statistical ensemble of realizations corresponding 
to a particular initial microstate ${\bf z}_A$ for the system
of interest is then defined operationally: it is the 
ensemble obtained by initializing $\psi$ in the microstate
${\bf z}_A$, then sampling the initial conditions of the
reservoirs (${\bf y}_1^0,\cdots,{\bf y}_N^0$)
from canonical distributions at the specified temperatures
($T_1,\cdots,T_N$).
Given ${\bf z}_A$, the choice of
$({\bf y}_1^0,\cdots,{\bf y}_N^0)$ uniquely
determines the subsequent evolution of all degrees of 
freedom, $\Gamma(t)$.
The probability distribution of obtaining a given realization,
conditional on a given microstate ${\bf z}_A$ for $\psi$,
then reduces to that of sampling the appropriate initial 
microstates of the reservoirs.

\section{Derivation}
\label{sec:derivation}

To carry out the derivation of Eq.\ref{eq:central},
we begin by introducing notation and spelling out
assumptions, starting with the classical approximation:
all quantal effects are ignored.

The system of interest, $\psi$, is taken to have a finite
number of degrees of freedom, and its instantaneous
microstate is described by a point
${\bf z} = ({\bf q},{\bf p})$
in the phase space of $\psi$, with the
usual assignment of ${\bf q}$ to denote configurational
variables, and ${\bf p}$ the associated momenta.
At any instant in time,
the internal energy of $\psi$
is given by a Hamiltonian $H_\lambda^\psi({\bf z})$,
parametrized by the current value of the
work parameter $\lambda$.

Next, assume that each heat reservoir $\theta_n$ 
is itself a classical system with 
a finite number of degrees of freedom, 
whose microstate is described by a point ${\bf y}_n$ 
in the phase space associated with that reservoir.
We do {\it not} assume the reservoirs to be physically
identical, so the dimensionalities of the
${\bf y}_n$'s may differ.
The internal energy of the $n$'th reservoir is given by
a Hamiltonian $H_n^\theta({\bf y}_n)$.

Finally, let $h_n^{\rm int}({\bf z},{\bf y}_n)$ denote
a weak coupling term between the system of interest 
and the $n$'th reservoir.
As discussed below, thermal contact between $\psi$ 
and $\theta_n$ can be established
or broken by turning this term ``on'' and ``off'',
using parameters $c_n(t)$.

For simplicity of presentation,
assume time-reversal invariance (``no magnetic fields'')
for these Hamiltonian terms:
\begin{equation}
\label{eq:tri}
H_\lambda^\psi({\bf z}^*) = 
H_\lambda^\psi({\bf z}) \quad,\quad
H_n^\theta({\bf y}_n^*)=H_n^\theta({\bf y}_n) \quad,\quad
h_n^{\rm int}({\bf z}^*,{\bf y}_n^*)= 
h_n^{\rm int}({\bf z},{\bf y}_n).
\end{equation}

The vector
\begin{equation}
\Gamma
= ( {\bf z},{\bf Y})
= ( {\bf z},{\bf y}_1,\cdots,{\bf y}_N)
\end{equation}
specifies the instantaneous state of all degrees
of freedom involved,
where ${\bf Y}=({\bf y}_1,\cdots,{\bf y}_N)$ 
denotes the collective microstate of the $N$ reservoirs.
The evolution of $\Gamma(t)$ is taken to be deterministic, 
and governed by a Hamiltonian
\begin{equation}
\label{eq:fullham}
{\cal H}(\Gamma,t) = 
H_{\lambda(t)}^\psi({\bf z}) +
\sum_{n=1}^N H_n^\theta({\bf y}_n) +
\sum_{n=1}^N c_n(t) \,
h_n^{\rm int}({\bf z},{\bf y}_n).
\end{equation}
Here $\lambda(t)$ denotes the time-dependence of the
work parameter, and the $c_n(t)$'s take on values of 0 or 1, 
which can also change with time.
At a given time $t$, if
$c_n(t)=1$, then this indicates that the system of 
interest is in thermal contact with the $n$'th reservoir 
at that time;
when $c_n(t)=0$, $\psi$ and $\theta_n$ are {\it not} in contact.
(More generally, we could let the $c_n$'s take on a
continuous range of values, allowing the thermal contacts
to be turned on and off smoothly rather than abruptly.
This modification would have no effect on the analysis.)

The collection 
$\{\lambda,\vec c\,\}\equiv\{\lambda,c_1,\cdots,c_N\}$
represents the ``tool-kit'' available for externally
manipulating the system of interest.
The protocol for a given process, $\Pi$, is then
just a particular prescription for doing so: it is
synonymous with a specific set of functions of time,
$\{\lambda(t),\vec c\,(t)\}$,
instructing us exactly how to manipulate the work
parameter and thermal contacts over a time
interval $[0,\tau]$.
This protocol uniquely specifies the 
time-dependence of the Hamiltonian function 
${\cal H}(\Gamma,t)$,
since that time-dependence enters only through
$\lambda(t)$ and $\vec c\,(t)$.
We will use
$ {\cal H}_\Pi (\Gamma,t)$ to denote
the time-dependent Hamiltonian corresponding to a
particular process $\Pi$.
If the protocol for a process $\Pi^+$
is $\{\lambda(t),\vec c\,(t)\}$, then that of its
time-reversed counterpart $\Pi^-$ is given by
$\{\lambda(\tau-t),\vec c\,(\tau-t)\}$.

While the time-dependence of the parameters 
$\{\lambda,\vec c\}$ is controlled externally,
the participating {\it dynamical} degrees of 
freedom $\Gamma=({\bf z},{\bf y}_1,\cdots,{\bf y}_N)$
evolve under Hamilton's equations, as
determined by ${\cal H}_ \Pi (\Gamma,t)$.
Thus, initial conditions $\Gamma(0)$
uniquely determine a trajectory $\Gamma(t)$,
which chronicles the microscopic history of all degrees
of freedom.
From this trajectory, as mentioned earlier, we can
extract both the microscopic history of the system
of interest, ${\bf z}(t)$ (by projecting out the reservoir 
variables ${\bf Y}$), and the net heat absorbed by 
$\psi$ from each of the reservoirs.
The heat absorbed by $\psi$ from a particular reservoir
is equal to the net decrease in the internal
energy of that reservoir:
\begin{equation}
\label{eq:defdq}
\Delta Q_n = H_n^\theta({\bf y}_n(0)) -
H_n^\theta({\bf y}_n(\tau)).
\end{equation}
From the $\Delta Q_n$'s we in turn compute
the entropy generated (Eq.\ref{eq:ent_gen}).

We note that, if $\Gamma_+(t)$ is a microscopic
realization of a process $\Pi^+$, then
$\Gamma_-(t)\equiv\Gamma_+^*(\tau-t)$
is a realization of the reverse process $\Pi^-$.
This follows from the assumption of time-reversal
invariance:
if $\Gamma_+(t)$ satisfies Hamilton's equations
for the forward process, then $\Gamma_-(t)$ will do
so for the reverse.

The reservoirs, as mentioned, are initially prepared
at specified temperatures, $T_1,\cdots,T_N$.
We take this to imply that their initial microstates
${\bf y}_n^0\equiv{\bf y}_n(0)$
are sampled from canonical ensembles.
This defines the following probability distribution
for the collection of initial reservoir conditions:
\begin{equation}
\label{eq:canon}
p({\bf Y}^0)=
{\cal N}^{-1} \prod_{n=1}^N
\exp \Bigl[-H_n^\theta({\bf y}_n^0)/k_BT_n\Bigr],
\end{equation}
where ${\cal N}(T_1,\cdots,T_N)$ is a
product of partition functions.

Finally, for a process $\Pi^+$ and a set of initial
conditions $\Gamma$ in the full phase space, let
\begin{equation}
\hat\Gamma_+^t(\Gamma) \equiv
\Bigl(\hat{\bf z}_+^t(\Gamma),\hat{\bf Y}_+^t(\Gamma)\Bigr)
\end{equation}
denote the point in phase space reached after time $t$,
and let $\Delta\hat S_+(\Gamma)$ denote the net
entropy generated over the entire realization of
the process (from $t=0$ to $t=\tau$).
The carats emphasize that
$\hat\Gamma_+^t$, $\hat{\bf z}_+^t$, $\hat{\bf Y}_+^t$, 
and $\Delta\hat S_+$ are viewed as {\it functions}
of the initial conditions $\Gamma$.
For the time-reversed process, we adopt the same notation,
with an obvious change in subscript
($\hat\Gamma_-^t$, $\hat{\bf z}_-^t$, etc.)

All the pieces needed to derive Eq.\ref{eq:central}
are now in place.
We begin with a formal expression for the joint,
conditional probability distribution in which we are
interested:
\begin{equation}
\label{eq:formal}
P_+({\bf z}_B,\Delta S\vert {\bf z}_A) =
\int d{\bf Y} p({\bf Y})
\delta[{\bf z}_B-\hat{\bf z}_+^\tau({\bf z}_A,{\bf Y})]
\cdot
\delta[\Delta S-\Delta\hat S_+({\bf z}_A,{\bf Y})].
\end{equation}
Using the identity
$\hat{\bf z}_+^0({\bf z},{\bf Y})={\bf z}$,
we rewrite this as:
\begin{equation}
P_+({\bf z}_B,\Delta S\vert {\bf z}_A) =
\int d\Gamma p({\bf Y})\cdot
\delta[{\bf z}_A-\hat{\bf z}_+^0(\Gamma)]\cdot
\delta[{\bf z}_B-\hat{\bf z}_+^\tau(\Gamma)]\cdot
\delta[\Delta S-\Delta\hat S_+(\Gamma)],
\end{equation}
where $\Gamma\equiv({\bf z},{\bf Y})$.
Now let 
$\Gamma^\prime=({\bf z}^\prime,{\bf Y}^\prime)=
\hat\Gamma_+^\tau(\Gamma)$
denote the {\it final} point in the full phase space,
for a realization of $\Pi^+$ launched from initial
conditions $\Gamma$.
For a given $\Gamma=({\bf z},{\bf Y})$,
we can rewrite $p({\bf Y})$ as:
\begin{equation}
p({\bf Y}) = 
{p({\bf Y})\over p({\bf Y}^\prime)}
p({\bf Y}^\prime) =
\exp\Bigl[
\Delta\hat S_+(\Gamma)/k_B\Bigr]
p({\bf Y}^\prime),
\end{equation}
using Eqs.\ref{eq:ent_gen}, \ref{eq:defdq}, \ref{eq:canon}, 
which leads to
\begin{equation}
\label{eq:interm}
P_+({\bf z}_B,\Delta S\vert {\bf z}_A) =
e^{\Delta S/k_B}
\int d\Gamma p({\bf Y}^\prime)
\delta[{\bf z}_A-\hat{\bf z}_+^0(\Gamma)]\cdot
\delta[{\bf z}_B-\hat{\bf z}_+^\tau(\Gamma)]\cdot
\delta[\Delta S-\Delta\hat S_+(\Gamma)].
\end{equation}
Here $p({\bf Y}^\prime)$ is {\it not} to be
interpreted as ``the probability distribution of
final reservoir conditions'', but rather as the
function $p$ defined by Eq.\ref{eq:canon}, evaluated
at ${\bf Y}^\prime=\hat{\bf Y}_+^\tau(\Gamma)$.
Since $\Gamma^\prime$ is reached from $\Gamma$ by time
evolution under the process $\Pi^+$, and since
we have assumed time-reversal invariance (Eq.\ref{eq:tri}),
it follows that, if we reverse the final momenta
and launch a realization of $\Pi^-$ from initial
conditions $\Gamma^{\prime *}$, then 
we will obtain the time-reversed image of the original
realization:
$\hat\Gamma_-^t(\Gamma^{\prime *}) =
[\hat\Gamma_+^{\tau-t}(\Gamma)]^*$.
From this it follows that
\begin{equation}
\hat{\bf z}_-^t(\Gamma^{\prime *}) =
\Bigl[ \hat{\bf z}_+^{\tau-t}(\Gamma) \Bigr]^* 
\quad,\quad
\Delta\hat S_-(\Gamma^{\prime *}) =
-\Delta\hat S_+(\Gamma).
\end{equation}
This allows us to rewrite Eq.\ref{eq:interm} as:
\begin{equation}
\label{eq:interm2}
P_+({\bf z}_B,\Delta S\vert {\bf z}_A) =
e^{\Delta S/k_B}
\int d\Gamma
p({\bf Y}^{\prime *})
\delta[{\bf z}_A^*-\hat{\bf z}_-^\tau(\Gamma^{\prime *})]\cdot
\delta[{\bf z}_B^*-\hat{\bf z}_-^0(\Gamma^{\prime *})]\cdot
\delta[\Delta S+\Delta\hat S_-(\Gamma^{\prime *})],
\end{equation}
where we have used the fact that 
$p({\bf Y}^\prime)=p({\bf Y}^{\prime *})$
(Eqs.\ref{eq:tri},\ref{eq:canon}).
Finally, since the integrand is expressed in terms of
$\Gamma^{\prime *}$, which is an invertible function of 
$\Gamma$ (defined by time evolution, followed by a reversal
of momenta), let us change the variables of integration 
from $\Gamma$ to $\Gamma^{\prime *}$.
The Jacobian for this change of variables is unity
(by Liouville's theorem),
so we simply replace $d\Gamma$ by
$d\Gamma^{\prime *}$ in Eq.\ref{eq:interm2}.
But then we can drop the prime and asterisk altogether
(since $\Gamma^{\prime *}$ is just a variable of
integration) to get:
\begin{eqnarray}
P_+({\bf z}_B,\Delta S\vert {\bf z}_A) &=&
e^{\Delta S/k_B}
\int d\Gamma p({\bf Y})
\delta[{\bf z}_A^*-\hat{\bf z}_-^\tau(\Gamma)]\cdot
\delta[{\bf z}_B^*-\hat{\bf z}_-^0(\Gamma)]\cdot
\delta[\Delta S+\Delta\hat S_-(\Gamma)] \\
&=&
e^{\Delta S/k_B}
\int d{\bf Y} p({\bf Y})
\delta[{\bf z}_A^*-\hat{\bf z}_-^\tau({\bf z}_B^*,{\bf Y})]\cdot
\delta[\Delta S+\Delta\hat S_-({\bf z}_B^*,{\bf Y})] \\
&=&
\label{eq:final}
e^{\Delta S/k_B}
P_-({\bf z}_A^*,-\Delta S\vert {\bf z}_B^*),
\end{eqnarray}
which is the desired result.

The origin of the exponential term
in Eq.\ref{eq:central} can be understood 
informally, as follows.
Given a ``forward'' realization
$\Gamma_+(t)$, and its time-reversed image
$\Gamma_-(t)$, $e^{\Delta S/k_B}$ 
is the probability
distribution for sampling the reservoir initial conditions
corresponding to the forward realization, relative to
those corresponding to the reverse realization, from
canonical distributions:
$e^{\Delta S/k_B}=p({\bf Y})/p({\bf Y}^{\prime *})$.
The probability $P_+$ appearing in the numerator
of Eq.\ref{eq:central}
is a sum of contributions from all
realizations for which $\psi$ evolves from
${\bf z}_A$ to ${\bf z}_B$
and the entropy generated is $+\Delta S$;
and similarly for $P_-$.
The two sets of realizations are in one-to-one
correspondence with each other:
for every $\Gamma_+(t)$ contributing to $P_+$
there is a time-reversed realization $\Gamma_-(t)$ 
contributing to $P_-$.
Since, for every such pair of realizations, the ratio
of probability distributions for sampling the associated 
initial reservoir conditions is $e^{\Delta S/k_B}$,
the ratio of the two sums
($P_+$ to $P_-$) is equal to this exponential.

We end this section by pointing out that a result
similar to Eq.\ref{eq:central} can be derived for
the statistic
\begin{equation}
P({\bf z}_1,{\bf z}_2,\cdots,{\bf z}_M,\Delta S\vert{\bf z}_0)
\qquad,\qquad M\ge 1,
\end{equation}
which is the joint probability distribution that $\psi$
will evolve through the sequence of points
${\bf z}_1,{\bf z}_2,\cdots,{\bf z}_M$, at times
$t_1,t_2,\cdots,t_M$, where $t_m=m\tau/M$,
and that the entropy generated will be $\Delta S$,
given ${\bf z}(0)={\bf z}_0$.
Formally, for a process $\Pi^+$,
\begin{equation}
P_+({\bf z}_1\cdots{\bf z}_M,\Delta S\vert{\bf z}_0) =
\int d{\bf Y} \,p({\bf Y})\,
\delta[\Delta S-\Delta\hat S_+({\bf z}_0,{\bf Y})]
\prod_{m=1}^M
\delta[{\bf z}_m-\hat{\bf z}_+^{t_m}({\bf z}_0,{\bf Y})].
\end{equation}
A calculation similar to the one presented above then
gives:
\begin{equation}
\label{eq:M}
{P_+({\bf z}_1,{\bf z}_2\cdots{\bf z}_M,+\Delta S
\vert{\bf z}_0\,) \over
P_-({\bf z}_{M-1}^*\cdots{\bf z}_0*,-\Delta S\vert
{\bf z}_M^*)} = \exp (\Delta S/k_B).
\end{equation}
Note that the discrete trajectory implied in the
denominator
$({\bf z}_M^*\rightarrow\cdots\rightarrow{\bf z}_0^*)$
is the time-reversed image of the one in the numerator
$({\bf z}_0\rightarrow\cdots\rightarrow{\bf z}_M)$.
Eq.\ref{eq:central} is just a special case, $M=1$,
of Eq.\ref{eq:M}.
The latter remains valid as well in the opposite
limit, $M\rightarrow\infty$ (with $\tau$ fixed), in which
case the entire history of $\psi$ is specified.
We then write, in suggestive notation,
\begin{equation}
\label{eq:Minf}
{P_+[{\bf z}_+(t),+\Delta S\vert{\bf z}_+(0)] \over
 P_-[{\bf z}_-(t),-\Delta S\vert{\bf z}_-(0)]}
= \exp (\Delta S/k_B),
\end{equation}
where ${\bf z}_-(t)={\bf z}_+^*(\tau-t)$.

For an {\it isolated} system ($N=0$) perturbed
by external forces of finite duration, 
we have $\Delta S=0$, by definition.
Thus, by Eq.\ref{eq:M}, the conditional probability 
distribution of observing 
the (isolated) system evolve through a given sequence
of points during the process $\Pi^+$, is equal to that of 
observing it to pass through the time-reversed sequence 
during $\Pi^-$.
This probability distribution will be a product of
$\delta$-functions: either the unique trajectory
launched from ${\bf z}_0$ goes through the sequence 
${\bf z}_1,\cdots,{\bf z}_M$, or it does not.
In this case ($N=0$),
Eq.\ref{eq:Minf} is essentially equivalent to Eq.7 of Ref.\cite{bk1}.
(A technical point of difference is
that Bochkov and Kuzovlev consider the {\it unconditional} 
probability of observing a given realization, 
{\it assuming the system of interest begins in equilibrium},
for both the forward and the reverse realization;
the exponential factor which they obtain is a
ratio of probabilities of sampling 
microstates of $\psi$ itself from a given equilibrium
distribution.)

For a single heat reservoir ($N=1$), Eq.\ref{eq:M} is 
similar to Eq.9 of Ref.\cite{gavin1}.
The main difference is that in Crooks' formulation
the evolution of the system of interest is explicitly taken
to be a Markov process, occurring in discrete steps.
Here, by contrast, the microstates ${\bf z}_m$ represent
``snapshots'' of $\psi$ taken at equally-spaced time 
intervals during continuous-time evolution, and 
in general this sequence of states cannot be viewed 
as a Markov chain.

Evans and Searles\cite{es} have also derived the (transient) 
fluctuation theorem by comparing the probabilities of sampling 
initial conditions of pairs of finite-time trajectories, 
one the time-reversed image of the other.
In their approach, the system of interest evolves under
deterministic but non-Hamiltonian equations of motion, 
to model the presence of a heat reservoir.
They find, as above, that the probability measure of a given 
trajectory, relative to that of its time-reversed twin,
is the exponent of the entropy generated (where the latter
is identified with phase space contraction).

\section{Relation to the steady-state fluctuation theorem}
\label{sec:ft}

In terms of the {\it average entropy generation rate},
$\overline\sigma\equiv\Delta S/\tau$,
Eq.\ref{eq:central} can be rewritten as
\begin{equation}
\label{eq:equivalent}
{1\over\tau}\ln
{P_+({\bf z}_B,+\overline\sigma\tau\vert {\bf z}_A) \over
 P_-({\bf z}_A^*,-\overline\sigma\tau\vert {\bf z}_B^*) }
= +\overline\sigma/k_B,
\end{equation}
which is reminiscent of
the steady-state fluctuation theorem, 
Eq.\ref{eq:ft}.
However, the correspondence is not exact:
Eq.\ref{eq:ft} applies explicitly to a
nonequilibrium steady state, contains the limit
$\tau\rightarrow\infty$, and exhibits no dependence on
initial and final microstates of the system of interest,
all in contrast to Eq.\ref{eq:equivalent}.
In this section we pursue the relationship
between the detailed fluctuation theorem of this paper
and the steady-state fluctuation theorem.

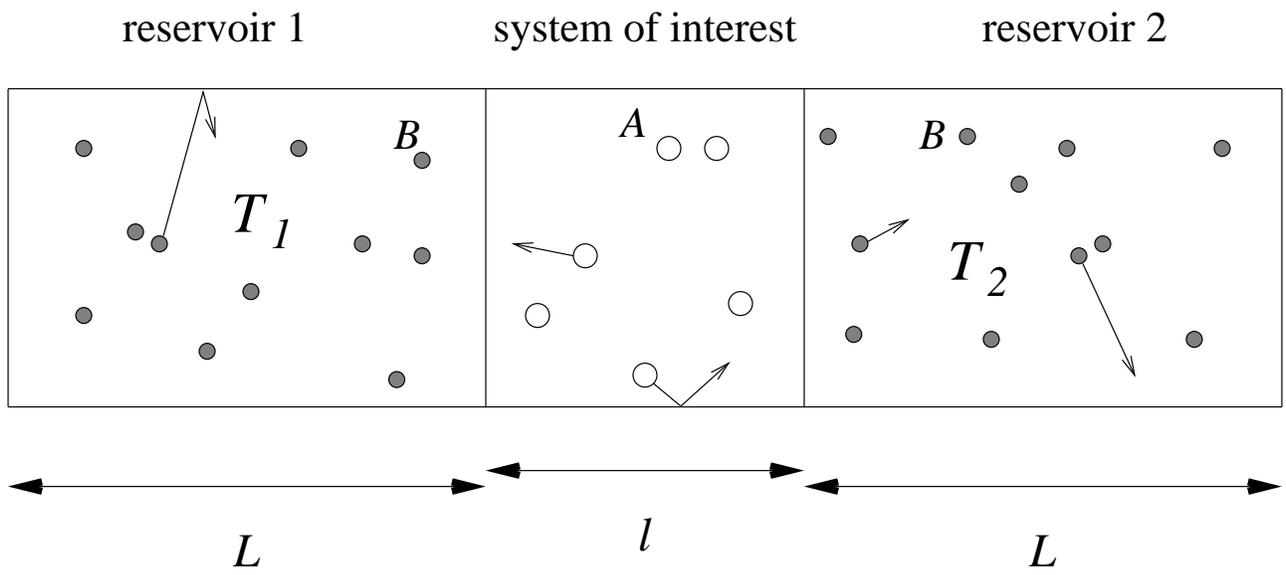
\begin{figure}
\caption{
Three interacting fluids.
See text for details.}
\label{fig:ness}
\end{figure}

Rather than aiming at complete generality, we will focus
on a specific physical situation which might exhibit a 
nonequilibrium steady state in the appropriate limit,
with the expectation that the line of reasoning applied 
here can serve as a model for other examples.
In Fig.\ref{fig:ness},
the system of interest is a fluid
composed of particles of ``type $A$'',  
inside a finite cylindrical container of length $l$.
The reservoirs are fluids of ``type $B$'' particles,
contained in two cylinders of length $L$
abutting the ends of $\psi$.
Let $\nu_\psi$ denote the number of particles
constituting the system of interest, and let 
$\nu_\theta=\nu_1=\nu_2$ denote the number in either reservoir.
Assume that the forces between particles are pairwise,
unaffected by the barriers between the cylinders,
and have a short interaction range $r<l$.
Also assume that all particles scatter elastically
off the container walls.

Given this set-up, one cannot expect the system to reach
a nonequilibrium steady state, {\it except possibly in
the limit of infinitely large reservoirs}.
We will now argue, quantitatively though not
rigorously, that if $\psi$ indeed
reaches a steady state in this limit, and if 
fluctuations in the entropy production in that
state are characterized by finite correlation times,
then Eq.\ref{eq:central} (or \ref{eq:equivalent})
implies the steady-state fluctuation theorem.

The system of interest and reservoirs depicted in
Fig.\ref{fig:ness} are governed by a Hamiltonian of the
form given in Eq.\ref{eq:fullham}, with
\begin{eqnarray}
H^\psi({\bf z}) &=&
\sum_{i=1}^{\nu_\psi}
{{\bf p}^{[i]2} \over 2m_A} +
\sum_{i<j} V_{AA}({\bf q}^{[i]},{\bf q}^{[j]}) + {\rm b.c.}\\
H_n^\theta({\bf y}_n) &=&
\sum_{i=1}^{\nu_\theta}
{{\bf p}_n^{[i]2} \over 2m_B} +
\sum_{i<j} V_{BB}({\bf q}_n^{[i]},{\bf q}_n^{[j]}) + {\rm b.c.}\\
h_n^{\rm int}({\bf z},{\bf y}_n) &=&
\sum_{i=1}^{\nu_\psi}
\sum_{j=1}^{\nu_\theta}
V_{AB}({\bf q}^{[i]},{\bf q}_n^{[j]}),
\end{eqnarray}
where the index $n=1,2$ labels the reservoirs.
Here, the microstates of the system of interest and
reservoirs are denoted by
\begin{eqnarray}
{\bf z} &=& ({\bf q}^{[1]},{\bf p}^{[1]},\cdots,
{\bf q}^{[\nu_\psi]},{\bf p}^{[\nu_\psi]})\\
{\bf y}_n &=& ({\bf q}_n^{[1]},{\bf p}_n^{[1]},\cdots,
{\bf q}_n^{[\nu_\theta]},{\bf p}_n^{[\nu_\theta]}),
\end{eqnarray}
$V_{xy}$ represents the short-range interaction 
potential between particles of type $x$ and $y$,
and ``b.c.'' denotes boundary conditions, implying
elastic reflection off the walls of the containers.
There is no work parameter, and thermal contact 
between the system of interest and the reservoirs is 
always ``on'' ($c_n=1$).

Let us now choose two temperatures $T_1$ and $T_2$
to be associated with the reservoirs $\theta_1$ and
$\theta_2$.
We can then subject the system of interest 
to a (static) thermodynamic process, by
starting with $\psi$ in some initial
microstate ${\bf z}_A$,
sampling the initial microstates
$({\bf y}_1^0,{\bf y}_2^0)$ of the reservoirs from 
canonical distributions at the chosen temperatures,
and letting the entire
system evolve for a time $\tau$ under the Hamiltonian
${\cal H} = H^\psi + \sum_n H_n^\theta + \sum_n h_n^{\rm int}$.
For this process, we can construct the statistic
$P_\tau^\omega({\bf z}_B,\Delta S\vert{\bf z}_A)$.
This is the joint, conditional probability
distribution defined in Section \ref{sec:central},
but with the dependence on the duration of the process,
$\tau$, and the size of the reservoirs,
$\omega\equiv(L,\nu_\theta)$, explicitly stated.

The detailed fluctuation theorem, 
Eq.\ref{eq:central}, then tells us that
\begin{equation}
\label{eq:Ptau}
{P_\tau^\omega({\bf z}_B,+\Delta S\vert{\bf z}_A) \over
P_\tau^\omega({\bf z}_A^*,-\Delta S\vert{\bf z}_B^*)} =
\exp (\Delta S/k_B),
\end{equation}
for this static process.
Let us now change variables,
from $\Delta S$ to $\overline\sigma=\Delta S/\tau$,
by defining
\begin{equation}
p_\tau^\omega({\bf z}_B,\overline\sigma\vert{\bf z}_A)
\equiv P_\tau^\omega({\bf z}_B,\overline\sigma\tau\vert
{\bf z}_A)\cdot\tau,
\end{equation}
the joint probability
distribution of observing, after a time $\tau$, a final 
microstate ${\bf z}_B$, and an average 
entropy generation rate $\overline\sigma$, conditional 
on an initial microstate ${\bf z}_A$.

We view 
$p_\tau^\omega({\bf z}_B,\overline\sigma\vert{\bf z}_A)$
as a function of ${\bf z}_A$, ${\bf z}_B$,
and $\overline\sigma$, parametrized by the values of
$\tau$, $L$, and $\nu_\theta$.
Let us now {\it assume}, first, that
\begin{equation}
\label{eq:assump1}
p_\tau^\Omega({\bf z}_B,\overline\sigma\vert{\bf z}_A) \equiv
\lim_{\omega\rightarrow\infty}
p_\tau^\omega({\bf z}_B,\overline\sigma\vert{\bf z}_A)
\quad{\rm exists \,(pointwise)},
\end{equation}
where ``$\omega\rightarrow\infty$'' denotes the limit
$L,\nu_\theta\rightarrow\infty$, 
with the particle density
$\nu_\theta/L$ held fixed.
Thus, we assume that
the dynamics of $\psi$ converges to a well-defined
limit, as we let the reservoirs become infinitely large.
Eq.\ref{eq:Ptau} then implies that
\begin{equation}
\label{eq:ptau}
{1\over\tau}\ln
{p_\tau^\Omega({\bf z}_B,+\overline\sigma\vert{\bf z}_A) \over
p_\tau^\Omega({\bf z}_A^*,-\overline\sigma\vert{\bf z}_B^*)}
={\overline\sigma\over k_B},
\end{equation}
for any finite value of $\tau$.

[A bit of care is needed here, since, in the limit
$\omega\rightarrow\infty$, the entropy generated becomes
defined in terms of differences between infinite
numbers (the initial and final energies of $\theta_1$
and $\theta_2$).
The assumption expressed by Eq.\ref{eq:assump1} states
that, for any {\it fixed} $\overline\sigma$,
the quantity 
$p_\tau^\omega({\bf z}_B,\overline\sigma\vert{\bf z}_A)$
converges to a particular value as the reservoirs
become increasingly larger;
roughly speaking, even though typical initial and
final reservoir energies diverge in that limit,
the energy differences $\Delta Q_n$ do not.
One can heuristically argue that this is a reasonable
assumption, as follows.
For a given particle density and temperature, there
ought to be a characteristic ``signal velocity'' $v$
with which information about the microstate of particles
in one region of the reservoir gets propagated to other
regions.
If we choose $L\gg v\tau$,
then we expect the particles at the
far end of either reservoir to have negligible influence
on the evolution of $\psi$, hence
$p_\tau^\omega({\bf z}_B,\overline\sigma\vert{\bf z}_A)$
will be unaffected by further increases in reservoir
size, $\omega$.
When taking the limit $\tau\rightarrow\infty$
below, it will be understood that the limit
$\omega\rightarrow\infty$ comes first.]

Let us next assume that, under the dynamics imposed by
infinitely large reservoirs, $\psi$
evolves to a statistical steady state. 
Let $f^S({\bf z})$ denote the distribution of
microstates, and $p_\tau^S(\overline\sigma)$
the probability distribution of observing an
average entropy production rate $\overline\sigma$
over a time interval of duration $\tau$, in the 
steady state.\footnote{
Note the distinction between the use of the subscript
$\tau$ in the statistic $p_\tau^S(\overline\sigma)$,
and its use in 
$p_\tau^\omega({\bf z}_B,\overline\sigma\vert{\bf z}_A)$.
In the latter, the time interval of duration $\tau$ is measured 
from the initial time, $t=0$, at which the reservoir
microstates are drawn from canonical distributions.
In the former, the interval is measured starting at
some moment {\it after the steady state has been achieved}.
In both cases, the entropy generated is defined in terms
of the net changes in internal energies of the reservoirs,
over the interval in question.}
We further assume that, in the steady state, fluctuations
in the entropy production are characterized by a finite
correlation time $t_c$, so that the average entropy
production rates measured over two adjacent time
intervals of duration $t_c$ can be treated as
statistically independent.
Finally, let $\overline\sigma^S$ denote the
infinite-time average entropy production rate
(equivalently, the {\it expectation value} of the average
entropy production rate over any finite time interval) 
in the steady state:
\begin{equation}
\lim_{\tau\rightarrow\infty}
p_\tau^S(\overline\sigma) =
\delta(\overline\sigma - \overline\sigma^S).
\end{equation}

Given these assumptions, we now want to justify replacing 
the numerator and denominator of Eq.\ref{eq:ptau} by
$p_\tau^S(+\overline\sigma)$ and $p_\tau^S(-\overline\sigma)$,
respectively, in the limit $\tau\rightarrow\infty$.

As a first step in this direction, we define
\begin{eqnarray}
p_\tau({\bf z}_B\vert{\bf z}_A) &\equiv&
\int d\overline\sigma\,
p_\tau^\Omega({\bf z}_B,\overline\sigma\vert{\bf z}_A)\\
{\rm and}\qquad
p_\tau(\overline\sigma\vert{\bf z}_A,{\bf z}_B)
&\equiv&
p_\tau^\Omega({\bf z}_B,\overline\sigma\vert{\bf z}_A)/
p_\tau({\bf z}_B\vert{\bf z}_A).
\end{eqnarray}
(We will drop the superscript $\Omega$ henceforth,
with the understanding that the
limit of infinite reservoirs is assumed in the remainder 
of this section.)
The former is the probability distribution of observing
a microstate ${\bf z}_B$ at $t=\tau$, given ${\bf z}_A$
at $t=0$.
The latter is the distribution of average entropy production
rates $\overline\sigma$ over the interval from
$t=0$ to $t=\tau$, conditional on an initial state
${\bf z}_A$ and a final state ${\bf z}_B$.
Note that
\begin{equation}
\label{eq:ssdist}
\lim_{\tau\rightarrow\infty}
p_\tau({\bf z}_B\vert{\bf z}_A) = f^S({\bf z}_B),
\end{equation}
by the assumption that $\psi$ evolves to a stationary
steady state.

Eq.\ref{eq:ptau} now becomes
\begin{equation}
\label{eq:sumlogs}
{1\over\tau}\ln
{p_\tau({\bf z}_B\vert{\bf z}_A) \over
 p_\tau({\bf z}_A^*\vert{\bf z}_B^*)} +
{1\over\tau}\ln
{p_\tau(+\overline\sigma\vert{\bf z}_A,{\bf z}_B) \over
 p_\tau(-\overline\sigma\vert{\bf z}_B^*,{\bf z}_A^*)} =
{\overline\sigma\over k_B},
\end{equation}
which again is valid for any $\tau$.
By Eq.\ref{eq:ssdist}, the first term on the left will 
vanish as $\tau\rightarrow\infty$.
It remains then to show that, in this limit, the numerator 
and denominator of the second term can be replaced by
$p_\tau^S(+\overline\sigma)$ and $p_\tau^S(-\overline\sigma)$.
It is tempting to argue that the dependence of 
$p_\tau(\overline\sigma\vert{\bf z}_A,{\bf z}_B)$
on the specified initial and final microstates will
vanish as $\tau\rightarrow\infty$, and therefore
\begin{equation}
\label{eq:limit}
p_\tau(\overline\sigma\vert{\bf z}_A,{\bf z}_B)
\rightarrow
p_\tau^S(\overline\sigma).
\end{equation}
This is true in the sense that both 
sides of Eq.\ref{eq:limit}
converge to the same 
distribution of values of $\overline\sigma$,
namely, $\delta(\overline\sigma-\overline\sigma^S)$.
However, since that limiting distribution is singular,
we cannot simply assume that, for instance, the ratio
of $p_\tau(\overline\sigma\vert{\bf z}_A,{\bf z}_B)$
to $p_\tau^S(\overline\sigma)$ at a given value of 
$\overline\sigma$
converges to unity as $\tau\rightarrow\infty$.
(In general it does not.)
Justifying the above-mentioned replacement will therefore
require some work.

Let $\underline x$ be a stochastic variable
denoting the time-averaged entropy production rate during an
interval of duration $t_c$, when the system is in the 
steady state.
Thus, the value of $\underline x$ is a value of
$\overline\sigma$ sampled randomly from the distribution
$p_{t_c}^S(\overline\sigma)$.
Note that $\langle\underline x\rangle=\overline\sigma^S$,
where angular brackets denote expectation value.
For an interval of duration $\tau=Kt_c$ (where $K$ is
a positive integer) we then have, by our assumption of
finite correlations,
\begin{equation}
\label{eq:stoch1}
p_\tau^S(\overline\sigma) = 
\langle\delta(\overline\sigma-\underline X)\rangle,
\end{equation}
where
\begin{equation}
\label{eq:X}
\underline X = {1\over K}\sum_{k=1}^K \underline x_k,
\end{equation}
and the $x_k$'s denote independent samples of the
same stochastic variable.

Eq.\ref{eq:stoch1} pertains to a system already in
the steady state.
Let us write down a similar equation for
$p_\tau(\overline\sigma\vert{\bf z}_A,{\bf z}_B)$.
For $\tau$ sufficiently large, we can divide the
interval $[0,\tau]$ into three segments:
initial, intermediate, and final.
During the initial segment, the system relaxes
to the steady state, and the influence of the initial
state ${\bf z}_A$ is felt statistically:
the probability distribution of values of entropy
produced during this segment depends on ${\bf z}_A$.
During the intermediate segment, the system is in the
steady state, and its behavior is independent of either 
${\bf z}_A$ or ${\bf z}_B$.
During the final segment, the influence of the
assumed final state ${\bf z}_B$ is felt statistically.
The amounts of entropy generated during each of these 
segments are statistically independent of one another.
For simplicity, let us assume that
the initial and final segments are both of duration
$t_c$, and that the total interval $\tau=Kt_c$ as
above.
Then we can write
\begin{equation}
p_\tau(\overline\sigma\vert{\bf z}_A,{\bf z}_B) =
\langle\delta(\overline\sigma-\underline Y)\rangle,
\end{equation}
where
\begin{equation}
\label{eq:Y}
\underline Y = {1\over K}
\Bigl(\underline a + \underline b +
\sum_{k=1}^{K-2} \underline x_k\Bigl),
\end{equation}
where $\underline a$ and $\underline b$ are
stochastic variables representing the average
entropy production rate during the initial
and final segments, respectively.
The dependence of these on ${\bf z}_A$ and ${\bf z}_B$
is implicit.

Both $p_\tau^S(\overline\sigma)$ and
$p_\tau(\overline\sigma\vert{\bf z}_A,{\bf z}_B)$
have been reduced to
distributions of averages of $K$
indpendently drawn samples.
In the former case, all $K$ samples are drawn from
the same distribution (Eq.\ref{eq:X}),
corresponding to the steady state.
In the latter case, $K-2$ samples are drawn from
that distribution, and the remaining two
($\underline a$ and $\underline b$) are drawn
otherwise (Eq.\ref{eq:Y}).
How do these two distributions compare, in the limit
$K\rightarrow\infty$ (in which one would expect the
contributions of $\underline a$ and $\underline b$
in Eq.\ref{eq:Y} to become negligible)?
Introducing 
$\underline c \equiv \underline a + \underline b$,
we can write
\begin{equation}
\label{eq:c}
p_\tau(\overline\sigma\vert{\bf z}_A,{\bf z}_B) = 
\int dc\,\eta(c)\,
p_{\tau^\prime}^S(\overline\sigma^\prime),
\end{equation}
where $\eta(c)=\langle\delta(c-\underline c)\rangle$
is the probability distribution of values
of $\underline c$; $\tau^\prime=\tau-2t_c$; and
\begin{equation}
\label{eq:sigmaprime}
\overline\sigma^\prime = {K\overline\sigma-c\over K-2}
\end{equation}
is the entropy generation rate implied
for the intermediate segment (of duration $\tau^\prime$), 
if the rates during the initial and final segments 
sum to a value $c$,
and the time-averaged rate for the entire interval
$[0,\tau]$ is $\overline\sigma$.
Thus, in Eq.\ref{eq:c} we are integrating over all possible 
ways of splitting a total entropy $\Delta S=\overline\sigma\tau$
into a sum of two terms: that generated during initial
and final segments ($ct_c$), and the remnant 
($\overline\sigma^\prime\tau^\prime$)
during the intermediate, steady-state segment.

The distribution of averages of many independently
drawn samples is governed by the
{\it theory of large deviations}\cite{lgdev},
which predicts that, for
a fixed value of $\overline\sigma$,
\begin{equation}
\label{eq:lgdev}
p_\tau^S(\overline\sigma)
=p_{Kt_c}^S(\overline\sigma)\rightarrow
\sqrt{ KI_0^{\prime\prime}\over\pi}
\exp\Bigl[-KI(\overline\sigma)\Bigr],
\end{equation}
as $\tau,K\rightarrow\infty$ ($t_c$ fixed).
Here, $I(x)$ is the {\it rate function}
(or {\it entropy function})
associated with the stochastic variable
$\underline x$;
$I_0^{\prime\prime}$ is the second
derivative of $I(x)$, evaluated at the minimum
of that function, $x_{\rm min}=\overline\sigma^S$;
and the normalization factor is obtained by steepest
descent.
[The rate function is defined up to an additive
constant.
For convenience we have set the value of $I$ to zero
at $x_{\rm min}$.
Eq.\ref{eq:lgdev} implies that, right around that minimum,
$p_\tau^S(\overline\sigma)$
tends toward a Gaussian of variance
$(2KI_0^{\prime\prime})^{-1}$.
This is just the central limit theorem.]
We can write down a similar result for
$p_{\tau^\prime}^S(\overline\sigma^\prime)$,
replacing $K$ by $K-2$ in Eq.\ref{eq:lgdev}, 
and $\overline\sigma$ by $\overline\sigma^\prime$.
Then taking the ratio of the two functions and
considering the limit $\tau\rightarrow\infty$
(equivalently, $K\rightarrow\infty$) gives
\begin{equation}
\label{eq:ratio}
\lim_{\tau\rightarrow\infty}
{p_{\tau^\prime}^S(\overline\sigma^\prime) \over
 p_\tau^S(\overline\sigma)} =
\exp\Bigl[
2I(\overline\sigma) + (c-2\overline\sigma)
I^\prime(\overline\sigma)\Bigr],
\end{equation}
where $I^\prime(x)\equiv dI(x)/dx$, and
$\overline\sigma^\prime$ is given by 
Eq.\ref{eq:sigmaprime}.

Combining Eqs.\ref{eq:c} and \ref{eq:ratio},
we finally get
\begin{equation}
\label{eq:lgdres}
\lim_{\tau\rightarrow\infty}
{p_\tau(\overline\sigma\vert{\bf z}_A,{\bf z}_B) \over
 p_\tau^S(\overline\sigma)} =
\int dc\,\eta(c)\,
\exp\Bigl[
2I(\overline\sigma) + (c-2\overline\sigma)
I^\prime(\overline\sigma)\Bigr]
\equiv R(\overline\sigma,{\bf z}_A,{\bf z}_B),
\end{equation}
where the dependence of $R$ on ${\bf z}_A$ and
${\bf z}_B$ enters through the implicit dependence of
$\underline a$ and $\underline b$ (hence, $\underline c$)
on those microstates
of $\psi$.
We see that, indeed, the ratio of the two distributions
does not generally converge to unity.
However, it {\it does} converge (by the arguments just 
presented, and assuming the integral in Eq.\ref{eq:lgdres}
converges!) to a function which does not depend on $\tau$.
Therefore we get, for the second term on the left side
of Eq.\ref{eq:sumlogs},
\begin{equation}
\label{eq:sumlogs2}
{1\over\tau}\ln
{p_\tau(+\overline\sigma\vert{\bf z}_A,{\bf z}_B) \over
 p_\tau(-\overline\sigma\vert{\bf z}_B^*,{\bf z}_A^*)}
\rightarrow
{1\over\tau}\ln
{p_\tau^S(+\overline\sigma) \over
p_\tau^S(-\overline\sigma)} +
{1\over\tau}\ln
{R(+\overline\sigma,{\bf z}_A,{\bf z}_B) \over
 R(-\overline\sigma,{\bf z}_B^*,{\bf z}_A^*)}
\end{equation}
as $\tau\rightarrow\infty$.
Combining Eqs.\ref{eq:sumlogs} and \ref{eq:sumlogs2}
and dropping the terms which vanish in that limit,
we finally get
\begin{equation}
\lim_{\tau\rightarrow\infty}
{1\over\tau}\ln
{p_\tau^S(+\overline\sigma)\over
p_\tau^S(-\overline\sigma)} = 
+\overline\sigma/k_B,
\end{equation}
which is the steady-state fluctuation theorem (Eq.\ref{eq:ft}).

While the arguments presented in this section
lack mathematical rigor,
they might point the way toward a proper derivation
of Eq.\ref{eq:ft} from Eq.\ref{eq:central}.
An interesting question is:
what additional assumptions are required in order for
the detailed fluctuation theorem
to rigorously imply the steady-state
fluctuation theorem?
Clearly a carefully crafted assumption about the
existence of a steady state in the limit of infinitely
large reservoirs ($\omega\rightarrow\infty$) 
is a {\it sine qua non}.
If we further assume exponential decay
of the autocorrelation function of the instantaneous
entropy generation rate,
then perhaps the arguments advanced
above could provide the backbone of a rigorous theorem.

A somewhat different approach has been taken by
Eckmann, Pillet, and Rey-Bellet\cite{epr},
in their study of a chain of anharmonic oscillators 
coupled to two infinite heat reservoirs.
They are able to project out the reservoir variables,
and thus reduce the evolution of the oscillator chain
(supplemented by a set of auxiliary variables) to
a Markov diffusion process.
By studying the generator of this diffusion process,
they argue that their model exhibits entropy
production in accordance with the steady-state
fluctuation theorem.

\section{Relation to far-from-equilibrium free energy results}
\label{sec:fe}

Independently of the fluctuation theorem, another
far-from-equilibrium result has been derived and
generalized in recent 
years\cite{cj1,cj2,gavin1,rn,acta,cd,hatano}.
Consider a system initially in thermal equilbrium
with a heat reservoir at temperature $T$.
Now imagine externally changing a work parameter
from an initial value (say, $\lambda=0$) to a 
final value ($\lambda=1$) over a finite time, while 
keeping the system in contact with the reservoir.
Once the final value of the work parameter has been
reached, hold the work parameter fixed
and let the system and reservoir re-equilibrate.
The system thus begins and ends in equilibrium states,
corresponding to $\lambda=0$ and $\lambda=1$,
but at intermediate times is driven out of equilibrium
by the finite-rate variation of the work parameter.
(The assumption that the system ends in equilibrium is not 
necessary, but makes for a more pleasant presentation.)
Now imagine repeating this process infinitely many
times, always following the same protocol for varying
$\lambda$, and carefully measuring the 
external {\it work} $W$ performed on the system during 
each realization.  
Then the distribution
of values of $W$ obtained from this ensemble of
realizations obeys the following equality, {\it regardless
of how gently or violently the parameter was switched}
from 0 to 1:
\begin{equation}
\label{eq:fed}
\Bigl\langle \exp (-\beta W)\Bigr\rangle =
\exp (-\beta\Delta F) \qquad,\qquad
\beta\equiv 1/k_BT,
\end{equation}
where $\langle\cdots\rangle$ denotes an average over the
ensemble of realizations, 
\begin{equation}
\Delta F = -\beta^{-1}\ln(Z_1/Z_0)
\end{equation}
is the free energy difference between the initial and
final equilibrium states of the system, and the $Z$'s
are the associated partition functions:
\begin{equation}
Z_\lambda = \int d{\bf z}
\exp[-\beta H_{\lambda}^\psi({\bf z})].
\end{equation}
Eq.\ref{eq:fed} was originally derived using a Hamiltonian
formulation\cite{cj1}, but has also been shown to be
valid under explicitly non-Hamiltonian evolution --
including the Nos\' e-Hoover thermostating scheme\cite{cj1,cj2},
Markov-chain dynamics\cite{cj2,gavin1,rn}, and Langevin
evolution\cite{cj2,acta} -- and has been 
generalized to a wider class of thermodynamic
processes\cite{gav2,gavin1,cd,hatano}.

[Eq.\ref{eq:fed} can be viewed as an extension,
to irreversible processes, of the relation
$W=\Delta F$, which holds for a {\it reversible},
isothermal process from one equilibrium state to
another.
Furthermore, it immediately implies the
inequality $\langle W\rangle\ge\Delta F$,
in agreement with the second law of thermodynamics,
and places an exponentially decaying upper bound 
on the probability of observing finite-size 
violations of the second law\cite{cd}:
\begin{equation}
{\rm Prob}(W\le\Delta F-X)\,\le\,e^{-X/k_BT}.]
\end{equation}

Let us now derive Eq.\ref{eq:fed} from the detailed
fluctuation theorem obtained in the present paper,
following a line of reasoning similar to that
presented by Crooks\cite{gavin1} for the case of
Markov evolution.

Let $\Pi^+$ be a process involving a system of interest,
$\psi$, a single reservoir, $\theta$ (prepared at
temperature $T$), and a work parameter $\lambda$ which
is varied from $\lambda(0)=0$ to $\lambda(\tau)=1$.
During a single realization of this process, the
{\it work} $W$ performed on $\psi$ is given by
\begin{equation}
W = \Delta E - \Delta Q,
\end{equation}
where
$\Delta E$ is the net change
in the internal energy of $\psi$, and $\Delta Q$
(Eq.\ref{eq:defdq}) is the heat absorbed by $\psi$
from the reservoir.
Following the notation of previous sections, let
${\bf z}_A$ and ${\bf z}_B$
denote the initial and final microstates of $\psi$,
and $\Delta S$ the entropy generated,
for a given realization.
Since $\Delta Q = -T\cdot\Delta S$ (Eq.\ref{eq:ent_gen}),
the value of $W$ can be expressed as a function of
${\bf z}_A$, ${\bf z}_B$, and $\Delta S$:
\begin{equation}
\label{eq:wzzs}
W({\bf z}_A,{\bf z}_B,\Delta S) =
H_{\lambda=1}^\psi({\bf z}_B) -
H_{\lambda=0}^\psi({\bf z}_A) + T\cdot\Delta S.
\end{equation}
Assuming a canonical distribution of initial conditions,
we can construct the ensemble average of $\exp(-\beta W)$
as follows:
\begin{equation}
\Bigl\langle\exp (-\beta W)\Bigr\rangle =
\int d{\bf z}_A 
{1\over Z_0} 
\exp[-\beta H_{\lambda=0}^\psi({\bf z}_A)]
\int d{\bf z}_B \int d(\Delta S)
P_+({\bf z}_B,\Delta S\vert{\bf z}_A)
\exp (-\beta W).
\end{equation}
Invoking Eqs.\ref{eq:central} and \ref{eq:wzzs} allows us 
to rewrite this as:
\begin{eqnarray}
\Bigl\langle\exp (-\beta W)\Bigr\rangle &=&
{1\over Z_0} \int d{\bf z}_A 
\exp [-\beta H_{\lambda=1}^\psi({\bf z}_B)] 
\int d{\bf z}_B \int d(\Delta S)
P_-({\bf z}_A^*,-\Delta S\vert{\bf z}_B^*)
\\
&=& {1\over Z_0} \int d{\bf z}_B
\exp [-\beta H_{\lambda=1}^\psi({\bf z}_B)] \\
&=& {Z_1\over Z_0} = \exp (-\beta\Delta F),
\end{eqnarray}
as promised.

Note that Eq.8 of Ref.\cite{bk1}
can be viewed as a special case of Eq.\ref{eq:fed}
above, for the situation in which the initial and final 
Hamiltonian functions are the same:
$H_{\lambda=0}({\bf z})=H_{\lambda=1}({\bf z})$,
hence the free energy difference is identically zero: 
$\Delta F=0$.

\section{Discussion}
\label{sec:disc}

The motivation for this paper has been a belief that
one ought to be able to derive the fluctuation theorem
-- or something like it -- within the framework of
traditional statistical mechanics;
that is, by contemplating a system of interest interacting
with a thermal environment.
The central result of this exercise, Eq.\ref{eq:central},
is a detailed fluctuation theorem,
valid for finite times
and made without reference to a steady state.
Given the definitions and assumptions
made in this paper, Eq.\ref{eq:central} is identically
true.
We now briefly address a number of issues related to this
result.

The first involves the definition of $\Delta S$.
Specifying what is meant by the entropy generated
during a single realization of a given process
is inherently problematic,
since entropy, in statistical mechanics, is ordinarily
associated with an {\it ensemble} of microstates.
Eq.\ref{eq:ent_gen} must
therefore be regarded as constituting a particular
{\it choice} of definition of $\Delta S$.
This choice, however, is not entirely arbitrary, but
rather (as indicated in Section \ref{sec:central})
guided by macroscopic thermodynamics,
where the entropy increase of a thermal environment
is identified with the heat, per unit temperature,
absorbed by that environment.

There is another apparently troubling feature of the
definition of entropy generated:
to compute $\Delta S$, it seems we must know the
exact initial and final microstates of each reservoir
(Eq.\ref{eq:defdq}).
This is extremely unsatisfying, as it conflicts with
the usual notion of a thermal environment as a huge
collection of unmonitored (and uninteresting) degrees
of freedom ``out there''.
However, one can infer the $\Delta Q_n$ values by
monitoring the evolution of only those environmental
degrees of freedom which are at any moment interacting
(exchanging energy) with the system of interest.
Therefore if, due to short-range interaction forces,
the exchange of energy between $\psi$
and the $\theta$'s occurs {\it locally} --
say, at an interface -- then $\Delta S$
can be computed by knowing only what goes on in the
immediate vicinity of the system of interest
(e.g.\ within boundary layers of width $r$
in Fig.\ref{fig:ness}),
{\it without} explicit knowledge of initial and final
reservoir energies.
Thus, while Eq.\ref{eq:defdq} is indisputably useful
as a device in the derivation of the detailed
fluctuation theorem, that final result is really
a statement about $\psi$ and the heat fluxes into
and out of $\psi$, rather than one about $\psi$
and initial and final reservoir energies.

Finally, there is a bit of arbitrariness even in the
definition of the heat absorbed by a given reservoir
(Eq.\ref{eq:defdq}), owing to the small but finite
interaction energy $h_{\rm int}$.
As with Eq.\ref{eq:ent_gen}, this definition represents
a particular choice, but again this choice is consistent
with macroscopic thermodynamics.

The derivation presented
in Section \ref{sec:derivation} explicitly
assumes that the reservoir degrees of freedom are
initially sampled canonically (Eq.\ref{eq:canon}).
The result itself, however, might not depend as strongly
on this assumption as the derivation suggests.
For macroscopically large reservoirs,
Eq.\ref{eq:central} ought to remain valid,
at least to an excellent approximation, if the initial
reservoir conditions are sampled from {\it micro}canonical,
rather than canonical, distributions.
The argument for this is similar to the usual one
for the equivalence of microcanonical and canonical
averages in the thermodynamic limit, and goes roughly
as follows.
Imagine constructing a joint, conditional probability
distribution
\begin{equation}
\tilde P({\bf z}_B,\Delta S\vert{\bf z}_A),
\end{equation}
defined in the same way as
$P({\bf z}_B,\Delta S\vert{\bf z}_A)$,
but with reservoir initial conditions sampled from
microcanonical distributions.
In this case the temperatures $T_n$ appearing in the
definition of $\Delta S$ are the ``microcanonical
temperatures'' of the heat reservoirs:
\begin{equation}
(k_BT)^{-1} = {\partial\over\partial E}
\ln \int d{\bf y}\delta\Bigl[E-H^\theta({\bf y})\Bigr].
\end{equation}
$\tilde P$ depends on the set of initial reservoir
energies, just as $P$ depends
on the initial reservoir temperatures.
Explicitly, we can write
\begin{equation}
\tilde P = \tilde P^{\epsilon_1\cdots \epsilon_N}
({\bf z}_B,\Delta S\vert{\bf z}_A) \qquad,\qquad
P = P^{T_1\cdots T_N}
({\bf z}_B,\Delta S\vert{\bf z}_A),
\end{equation}
where $\epsilon_n$ denotes the known initial energy 
{\it per particle} (alternatively, per degree of freedom)
of the $n$'th reservoir: 
$\epsilon_n=E_n/\nu_n$, where $E_n$ is the initial
energy of, and $\nu_n$ the number of particles constituting,
the $n$'th reservoir.
Since the $\nu_n$'s are fixed,
the initial microcanonical ensemble of each reservoir
is uniquely specified by the value of $\epsilon_n$.
Now, $P^{T_1\cdots T_N}$ can be expressed as a
weighted average of $\tilde P^{\epsilon_1\cdots \epsilon_N}$:
\begin{equation}
P^{T_1\cdots T_N}
({\bf z}_B,\Delta S\vert{\bf z}_A) =
\Biggl[\prod_{n=1}^N \int d\epsilon_n \, w_n(\epsilon_n;T_n)\Biggr]\,
\tilde P^{\epsilon_1\cdots \epsilon_N}
({\bf z}_B,\Delta S\vert{\bf z}_A),
\end{equation}
where $w_n(\epsilon_n;T_n)$ is the statistical weight of
the microcanonical ensemble at energy-per-particle 
$\epsilon_n$,
within the canonical ensemble at temperature $T_n$,
for the $n$'th reservoir.\footnote{
Formally,
$w(\epsilon;T) \propto
e^{-\nu\epsilon/k_BT}
\int d{\bf y}\,\delta[\nu\epsilon-H^\theta({\bf y})]$,
with the subscript $n$ suppressed. 
The normalization is chosen so that $\int d\epsilon\,w=1$.}
In the thermodynamic limit of arbitrarily
large reservoirs ($\nu_n\rightarrow\infty$, with extensive
and intensive quantities scaled appropriately),
$w_n(\epsilon_n;T_n)$ becomes peaked arbitrarily sharply
around the value $\epsilon_n(T_n)$ whose
corresponding
``microcanonical temperature'' is equal to $T_n$.
Assuming as in Section \ref{sec:ft} that $\tilde P$
converges to a well-defined function of ${\bf z}_A$,
${\bf z}_B$, and $\Delta S$, we will therefore get, 
in the thermodynamic limit,
\begin{equation}
P^{T_1\cdots T_N}
({\bf z}_B,\Delta S\vert{\bf z}_A)= 
\tilde P^{\epsilon_1(T_1)\cdots \epsilon_N(T_N)}
({\bf z}_B,\Delta S\vert{\bf z}_A),
\end{equation}
resulting in a microcanonical detailed fluctuation
theorem.

More generally, we expect Eq.\ref{eq:central} to remain
valid, provided that each reservoir is prepared in what
would be characterized at the macroscopic level as a
state of thermodynamic equilibrium (regardless of whether
the method of preparation truly yields a canonical
distribution of microstates when carried out
repeatedly).
This reflects a strong prejudice that the canonical
ensemble should be viewed primarily as a computational
convenience, and ought not to be taken too seriously
as characterizing the ``true'' statistical distribution
of microstates of entire macroscopic bodies.
Of course, all this presupposes macroscopically large
heat reservoirs;
for microscopic ``reservoirs'', the detailed fluctuation
theorem still holds, but its validity then depends
on a literal interpretation of Eq.\ref{eq:canon}.

Another issue involves the assumption of time-reversal
invariance, Eq.\ref{eq:tri}.
While this assumption was made for convenience, it is
straightforward to generalize the analysis to the
situation in which (possibly time-dependent) magnetic 
fields are present.
In that case, given a process $\Pi^+$, its
time-reversed counterpart $\Pi^-$ is obtained by
carrying out the protocol in reverse order, while
also reversing all magnetic fields:
${\bf B}(t)\rightarrow-{\bf B}(\tau-t)$.
With this modification, Eq.\ref{eq:central} remains
valid (see also Refs.\cite{gav2,bk1,bk2}).

The derivation of the detailed fluctuation theorem presented 
in this paper is admittedly not as ``clean'' as in previous 
works\cite{ecm,es,gc,g1,c1,bgg,g2,g3,bcl,r,aes,se2,se3,gc2,k,ls,se1,gav2,maes},
where the (deterministic or stochastic) thermostating is
accomplished without the explicit introduction of reservoir
degrees of freedom.
For instance, the present treatment forces us to consider
the unphysical limit of infinite reservoirs;
caveats need to be made about ignoring interaction
energies; the time $t=0$ is privileged (since that is
when the reservoir microstates are sampled canonically);
and so forth.
There is some consolation, however, in knowing that
these messy issues are likely to arise in any laboratory
setting, where real thermal environments  and finite times
of observation are an inherent part of the game.

Nos\' e-Hoover-type thermostating schemes\cite{nh} --
employing one or a handful of ``reservoir'' degrees
of freedom (and non-Hamiltonian equations of motion) --
might offer an interesting middle ground between the
deterministic thermostats of 
Refs.\cite{ecm,es,gc,g1,c1,bgg,g2,g3,bcl,r,aes,se2,se3,gc2}
and the approach taken here.
In the original Nos\' e-Hoover scheme, the reservoir
degree of freedom ($\zeta$) is initially sampled 
from a Gaussian distribution.
Therefore, given an appropriate definition of
$\Delta S$, it would be straightforward to define a
joint, conditional probability distribution
$P({\bf z}_B,\Delta S\vert {\bf z}_A)$, as in this
paper, basically replacing the canonical distribution
of initial reservoir conditions in Eq.\ref{eq:formal}
by the Gaussian distribution of initial values of
$\zeta$.
It would be interesting to see whether a detailed
fluctuation theorem then follows.
If so, this approach might lead (by arguments along
the lines of Section \ref{sec:ft}) to a Nos\' e-Hoover
steady-state fluctuation theorem.
Presumably some chaoticity assumption would still be
required to make this result rigorous, but at least there 
would be no need to worry about infinitely large reservoirs!

Finally, it would be very nice to come up with a laboratory
experiment to test the main result derived in this paper.
The system of interest in such an experiment would
doubtless have to have very few degrees of freedom
(ideally, only one), in order to collect data with
sufficiently good statistics in a reasonable amount
of time.
Furthermore, one would want a process during which
the typical entropy generated is not much greater
than $k_B$;
otherwise, prohibitively many realizations would be
needed before observing a single one for
which $\Delta S<0$.
In this respect, the fact that Eq.\ref{eq:central}
is valid for {\it finite} durations $\tau$ is helpful.

\section*{Acknowledgments}
It is a pleasure to acknowledge and thank
F.Bonetto, G.E.Crooks, J.R.Dorfman, R.Mainieri, Y.Oono,
J.Percus, L.Rey-Bellet, and D.J.Searles
for stimulating conversations and correspondence.
This research is supported by the Department of Energy, 
under contract W-7405-ENG-36.

\end{document}